\documentclass{aastex62}

\begin{document}

\title{Deep Hyper Suprime-Cam Images and a Forced Photometry Catalog in W-CDF-S}

\correspondingauthor{Qingling Ni, John Timlin}
\email{qxn1@psu.edu, jxt811@psu.edu}

\newcommand\pennstate{Department of Astronomy \& Astrophysics, 525 Davey Lab, The Pennsylvania State University, University Park, PA 16802, USA}

\author{Q. Ni}
\affiliation{\pennstate}

\author{J. Timlin}
\affiliation{\pennstate}

\author{W. N. Brandt}
\affiliation{\pennstate}

\author{G. Yang}
\affiliation{\pennstate}

\begin{abstract}

The Wide Chandra Deep Field-South (W-CDF-S) field is one of the SERVS fields with extensive multiwavelength datasets, which can provide insights into the nature and properties of objects in this field.
However, the public optical data from DES (\textit{grizy} to $m\rm_{AB} \approx 21.4-24.3$) are not sufficiently deep to match well the NIR data from VIDEO ($ZYJHK_s$ to $m\rm_{AB} \approx 23.5-25.7$), which limits the investigation of fainter objects at higher redshifts.
Here, we present an optical catalog of $\approx$ 2,000,000 objects in W-CDF-S utilizing archival Hyper Suprime-Cam observations in the \textit{g,r,i,z} bands covering $\approx5.7 \rm~deg^2$. The estimated depth is $\approx$ 25.9 for \textit{g}-band, 25.6 for \textit{r}-band, 25.8 for \textit{i}-band, and 25.2 for \textit{z}-band, which is deep enough to complement the NIR data, and will benefit AGN/galaxy studies in W-CDF-S in the future.
\end{abstract}

\section{} 

The Wide Chandra Deep Field-South (W-CDF-S) region is a $\approx$ 4.5 deg$^2$ field in the SERVS \citep{Mauduit2012} CDF-S footprint. 
W-CDF-S has extensive multiwavelength coverage that is publicly available,
such as ATLAS \citep{Franzen2015} in the radio, HerMES \citep{Oliver2012} in the FIR, SERVS and VIDEO \citep{Jarvis2013} in the NIR, and DES \citep{Abbott2018} in the optical. PRIMUS \citep{Coil2011} also provides more than 30,000 spectroscopic redshifts in the field (see Table 1 of \citealt{Chen2018} for more details).
Furthermore, W-CDF-S will have $\approx$ 2.2 Ms of \textit{XMM-Newton} coverage by 2020 covering the whole field, in addition to the archival X-ray coverage in the $\approx$~0.3 deg$^2$ CDF-S \citep{Comastri2016,Xue2016,Luo2017}. 
However, the current public optical data from DES (\textit{grizy} to $m\rm_{AB} \approx 21.4-24.3$) do not delve as deep as the NIR data from VIDEO ($ZYJHK_s$ to $m\rm_{AB} \approx 23.5-25.7$). Thus, optical data at least as deep as the NIR data will benefit the investigation of fainter objects at higher redshifts.

\begin{figure}[h!]
\begin{center}
\includegraphics[scale=0.5,angle=0]{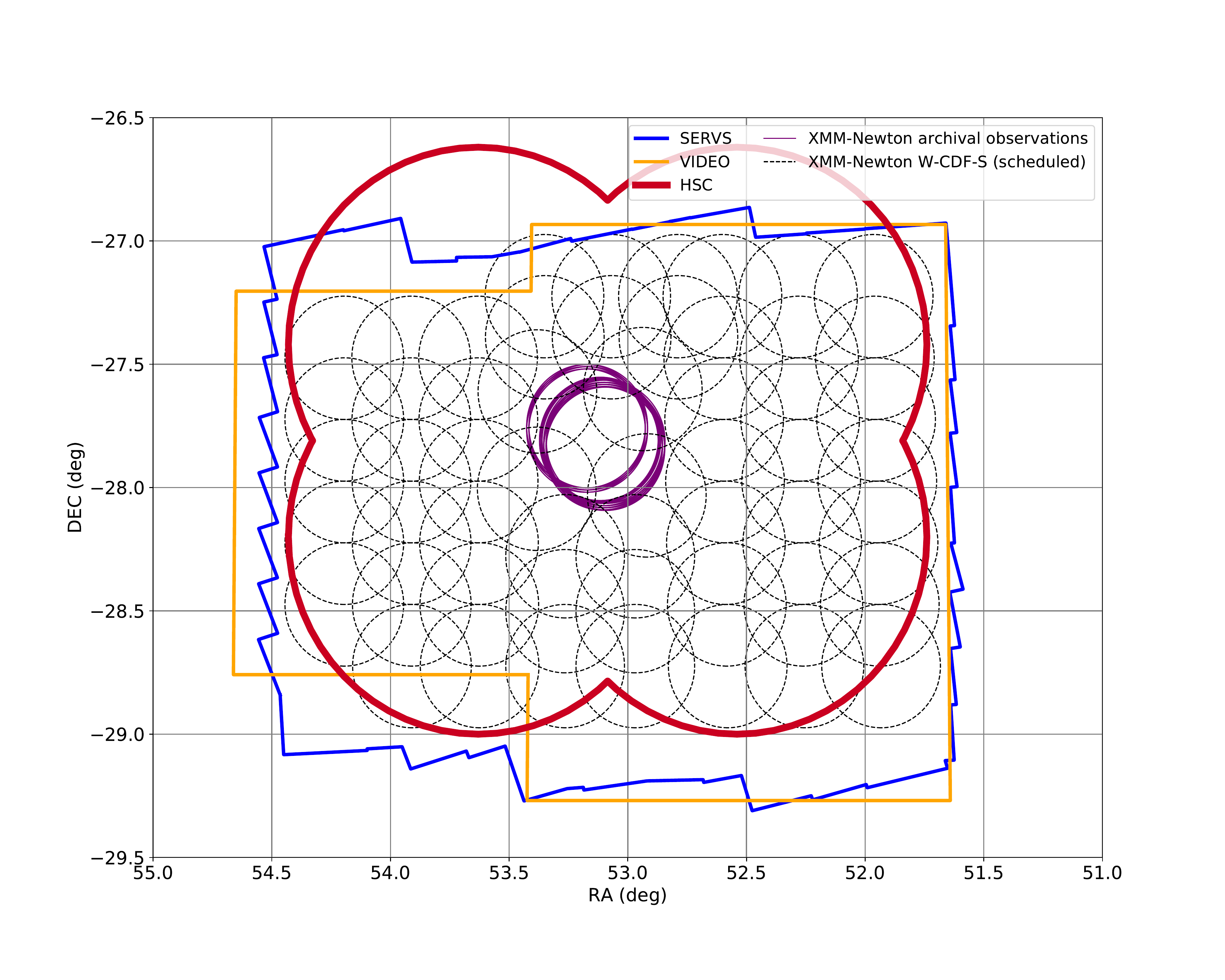}
\caption{
The multiwavelength coverage of the W-CDF-S field shown as labelled.\label{fig:1}}
\end{center}
\end{figure}

Here, we present an optical catalog of archival Hyper Suprime-Cam (HSC; \citealt{Miyazaki2012}) observations in W-CDF-S. 
HSC is an optical digital camera attached to the Subaru Telescope. 
We obtained the publicly available raw image data in the \textit{g,r,i,z} bands (taken between January 2015 and March 2017) and relevant calibration files via SMOKA.\footnote{\url{https://smoka.nao.ac.jp}} The coverage of HSC is shown in Figure \ref{fig:1} along with some other key multiwavelength data in the W-CDF-S region. Observations in the \textit{g,i,z} bands cover the full $\approx5.7 \rm~deg^2$ area shown in Figure \ref{fig:1}, while observations in the $r$-band only cover the central $\approx2.0 \rm~deg^2$ region. These observations have exposure times close to the Deep fields in the HSC Subaru Strategic Program \citep{Aihara2018}.

We analyzed the HSC data with \texttt{hscPipe} v5.4 \citep{Bosch2018}, a version of the LSST Software Stack \citep{Ivezic2008, Axelrod2010,Juric2017}, following the \texttt{hscPipe} user manual.\footnote{\url{https://hsc.mtk.nao.ac.jp/pipedoc_5_e/}} The pipeline first performs single-visit processing, which includes the subtraction of overscan, bias, and dark, and also flatfielding.
Then, it calibrates the relative position and flux scale of each CCD against the PanSTARRS1 (PS1) PV3 catalog \citep{Magnier2013}.
Utilizing the corrected position and flux scale, CCD images are warped and combined using a weighted average to reduce contamination.
Sources are then detected and measured in the \textit{g,r,i,z} bands separately, and a list of sources is generated by merging the object information from all the bands. Finally, forced photometry is performed in the four bands simultaneously with source positions and shape parameters fixed to the values of a well-detected reference band selected by \texttt{hscPipe} (see \citealt{Bosch2018} for details).

We present the coadded images and the forced photometry catalog at\dataset[10.5281/zenodo.2225161]{\doi{10.5281/zenodo.2225161}}. 
Corresponding weight maps and mask images are also provided, which contain information about the coadded image quality pixel by pixel. 
To use the catalog products, users should use the suggested flag cuts (e.g., Table~4 of \citealt{Aihara2018}) to select objects with clean photometry. We also created a \texttt{flag\_clean} column to mark ``clean'' objects (the selection criteria are shown in the catalog schema).

We also provide a few diagnostic plots to demonstrate the astrometric and photometric quality of the catalog in the Appendix. In Figure S1, we show the number of sources as a function of the PSF magnitude in each band for ``clean'' objects. The estimated depth is $\approx$ 25.9 for \textit{g}-band, 25.6 for \textit{r}-band, 25.8 for \textit{i}-band, and 25.2 for \textit{z}-band. 
The depths as a function of the patch location are shown in Figure S2. 
Also, HSC-to-$Gaia$ \citep{Gaia2018} positional offsets are presented in Figure S3, demonstrating the astrometric quality of the HSC data products.
Other diagnostic plots are also available.

Finally, we provide a preliminary match of the HSC sources to their infrared counterparts. We matched the ``clean'' HSC objects as well as detections in VIDEO to the SERVS source positions within 1$''$ to generate an optical-IR catalog of sources in W-CDF-S. 

We note that the VOICE Survey \citep{Vacarri2016} data will be released in the near future, which has $u,g,r,i$ bands down to $m\rm _{AB} \approx 25-26$ for a somewhat different 4 deg$^2$ footprint. Also, the ``Cosmic Dawn Survey'' will provide deeper \textit{Spitzer} (PI: P. Capak) and HSC (PIs: D. Sanders \& P. Capak) coverage in W-CDF-S in the coming years.

\acknowledgments
We thank the anonymous PI of the HSC data products we used in W-CDF-S.
We thank Yoshihiko Yamada at the \texttt{hscPipe} help desk, for assistance with \texttt{hscPipe}.
We thank Ian Smail and Michael Strauss for helpful discussions.

\appendix
\renewcommand\thefigure{S1} 
\begin{figure}[htbp]
   \centering
   \includegraphics[scale=0.7]{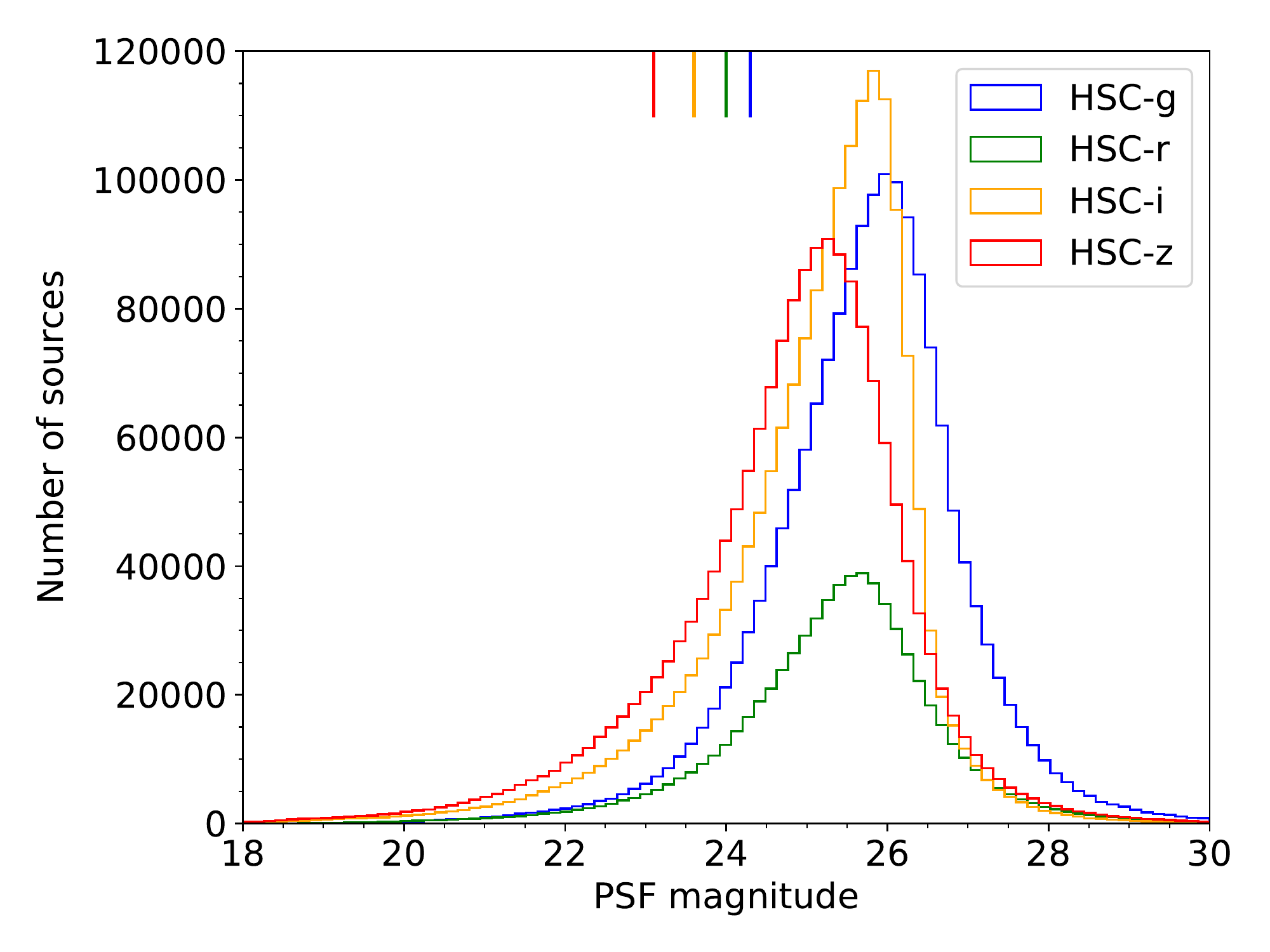}
   \caption{Histograms showing the distributions of PSF magnitudes for ``clean'' objects in the coadded HSC catalog. An estimation of the observation depth can be obtained by finding the mode of the magnitude distribution. The estimated depth is $\approx$ 25.9 for \textit{g}-band, 25.6 for \textit{r}-band, 25.8 for \textit{i}-band, and 25.2 for \textit{z}-band. For comparison, the depths of DES DR1 data in the $g,r,i,z$ bands are plotted as vertical bars with the color blue, green, orange, and red.}
\end{figure}

\renewcommand\thefigure{S2} 
\begin{figure}[htbp]
   \centering
   \includegraphics[scale=0.36]{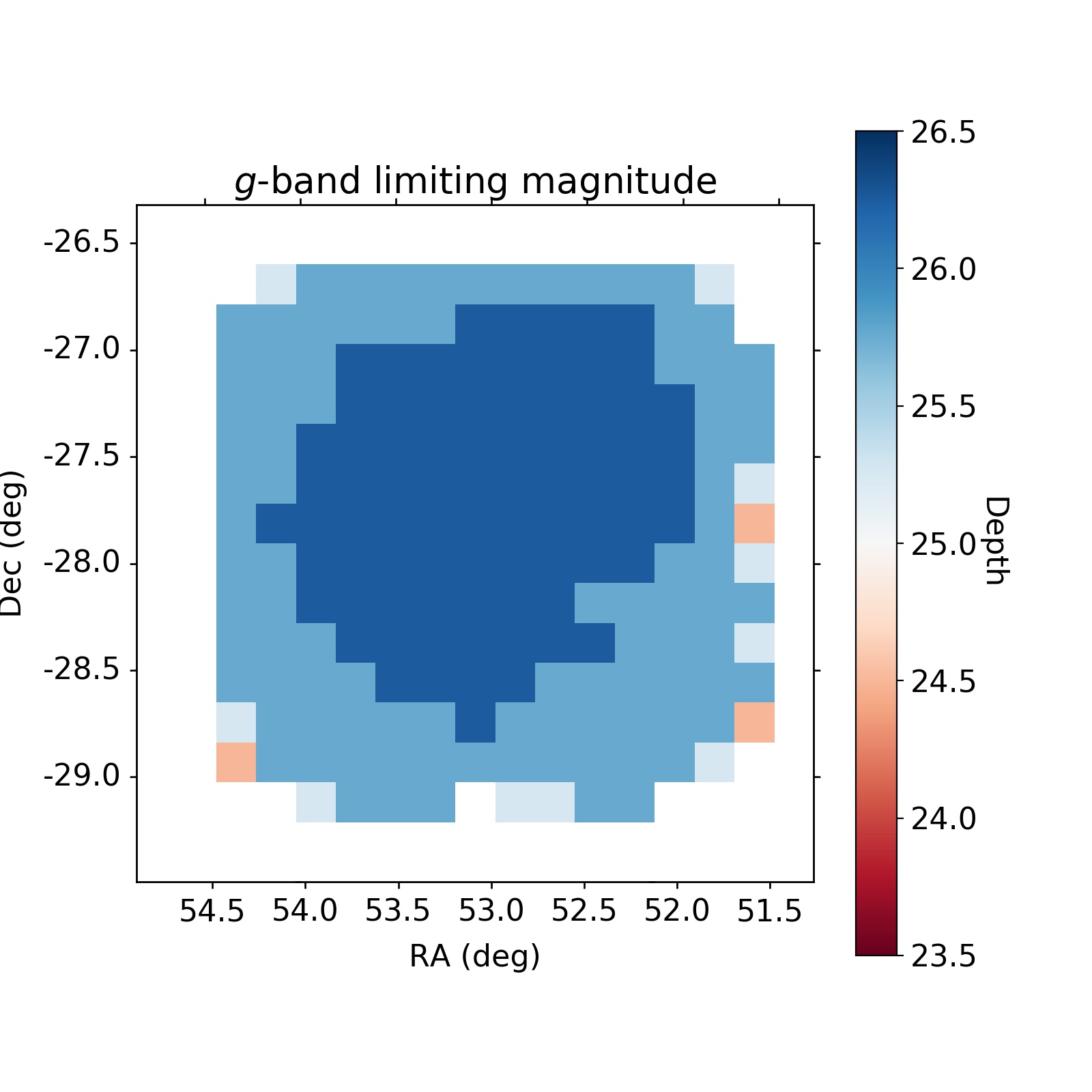}
   \includegraphics[scale=0.36]{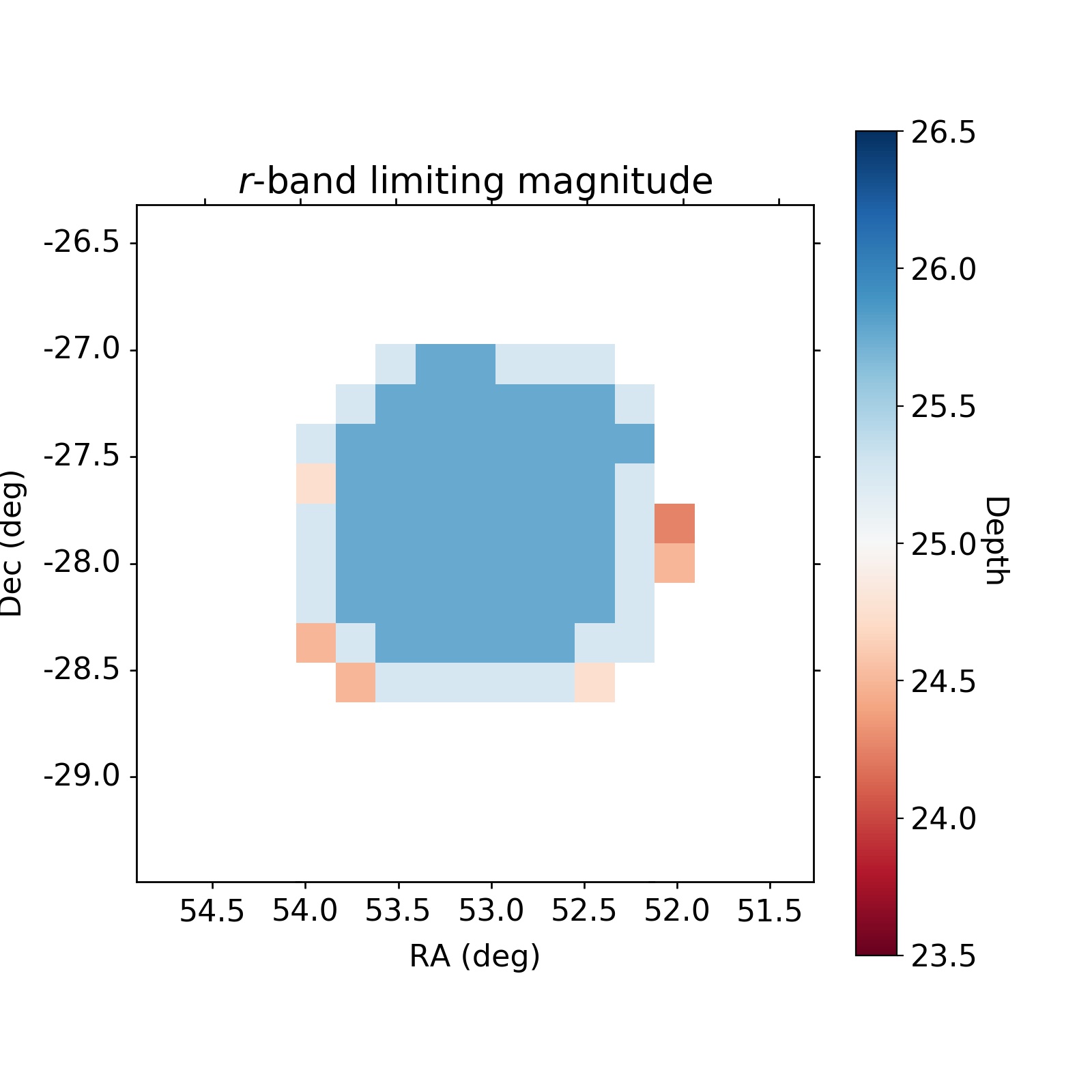}
   \includegraphics[scale=0.36]{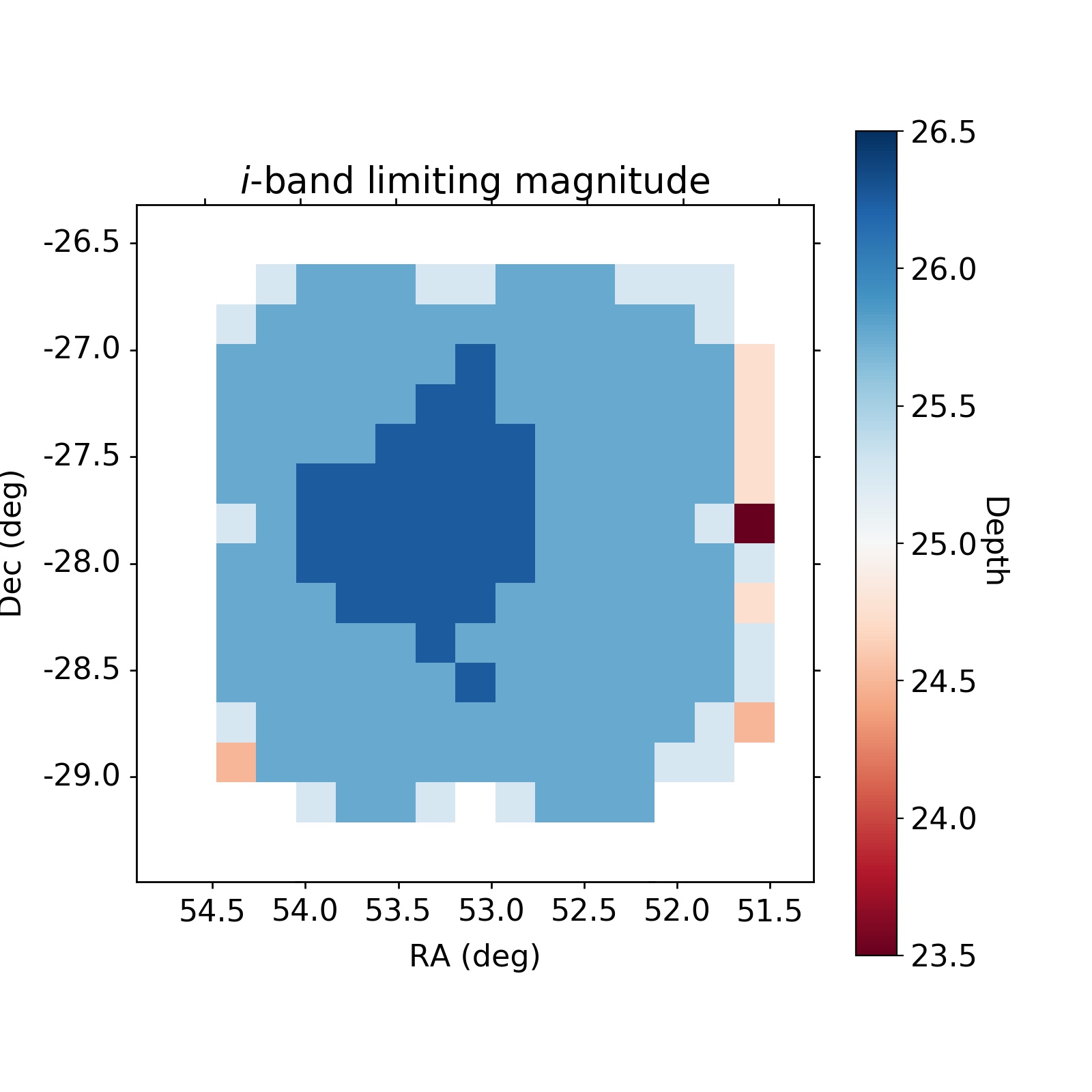}
   \includegraphics[scale=0.36]{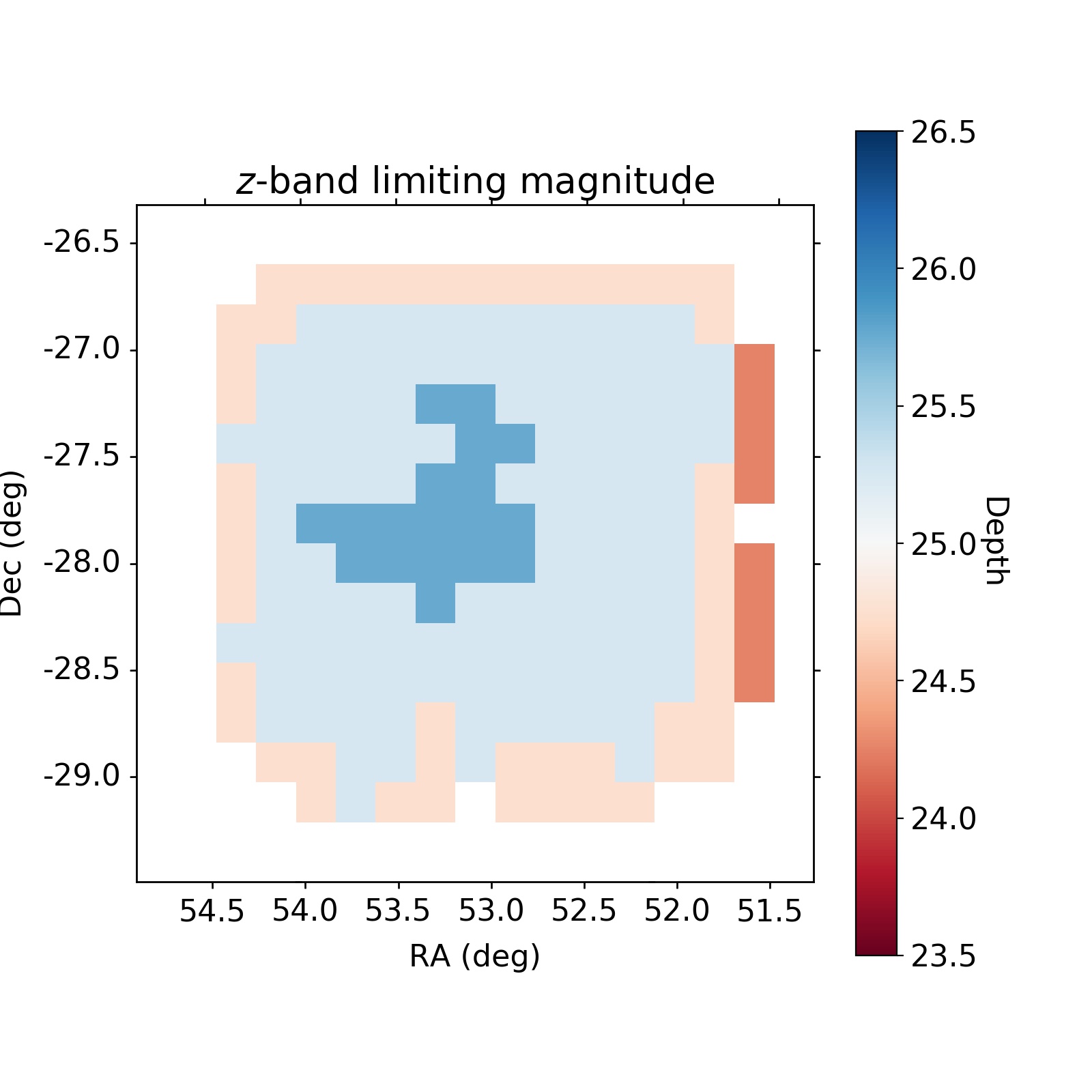}
   \caption{Patch-by-patch depth maps for the $g,r,i,z$ bands. The depths here are estimated by the mode of the PSF magnitude distribution in each patch.}
\end{figure}

\renewcommand\thefigure{S3}
\begin{figure}[htbp]
   \centering
   \includegraphics[scale=0.45]{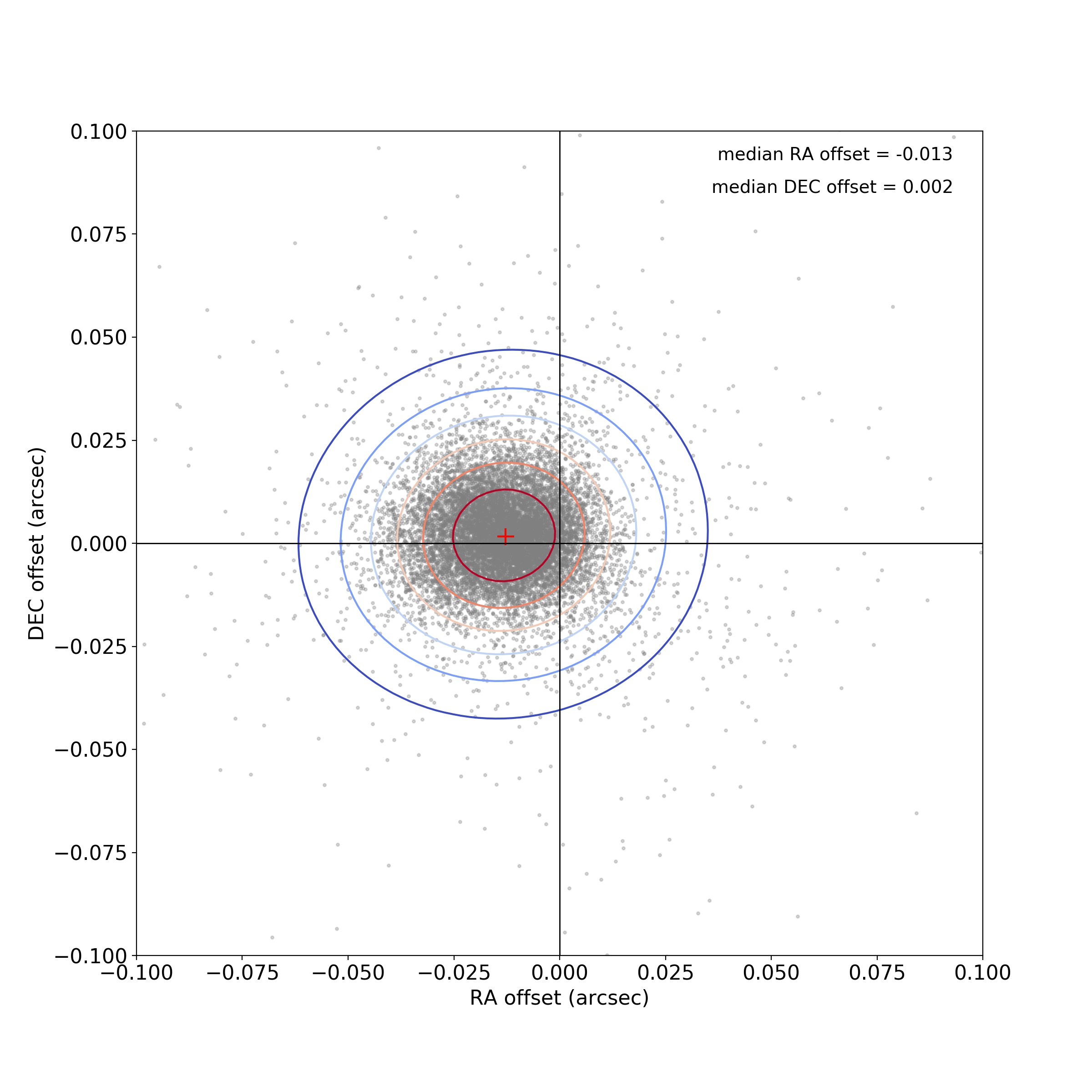}
   \caption{Distribution of the HSC-to-$Gaia$ positional offsets in the RA versus Dec plane. The contours represent the isodensity levels of the points. The median positional offset is represented by the red cross. We give equations to correct to the $Gaia$ astrometric frame in \texttt{readme.txt}.}
\end{figure}

\renewcommand\thefigure{S4} 
\begin{figure}[htbp]
   \centering
   \includegraphics[scale=0.28]{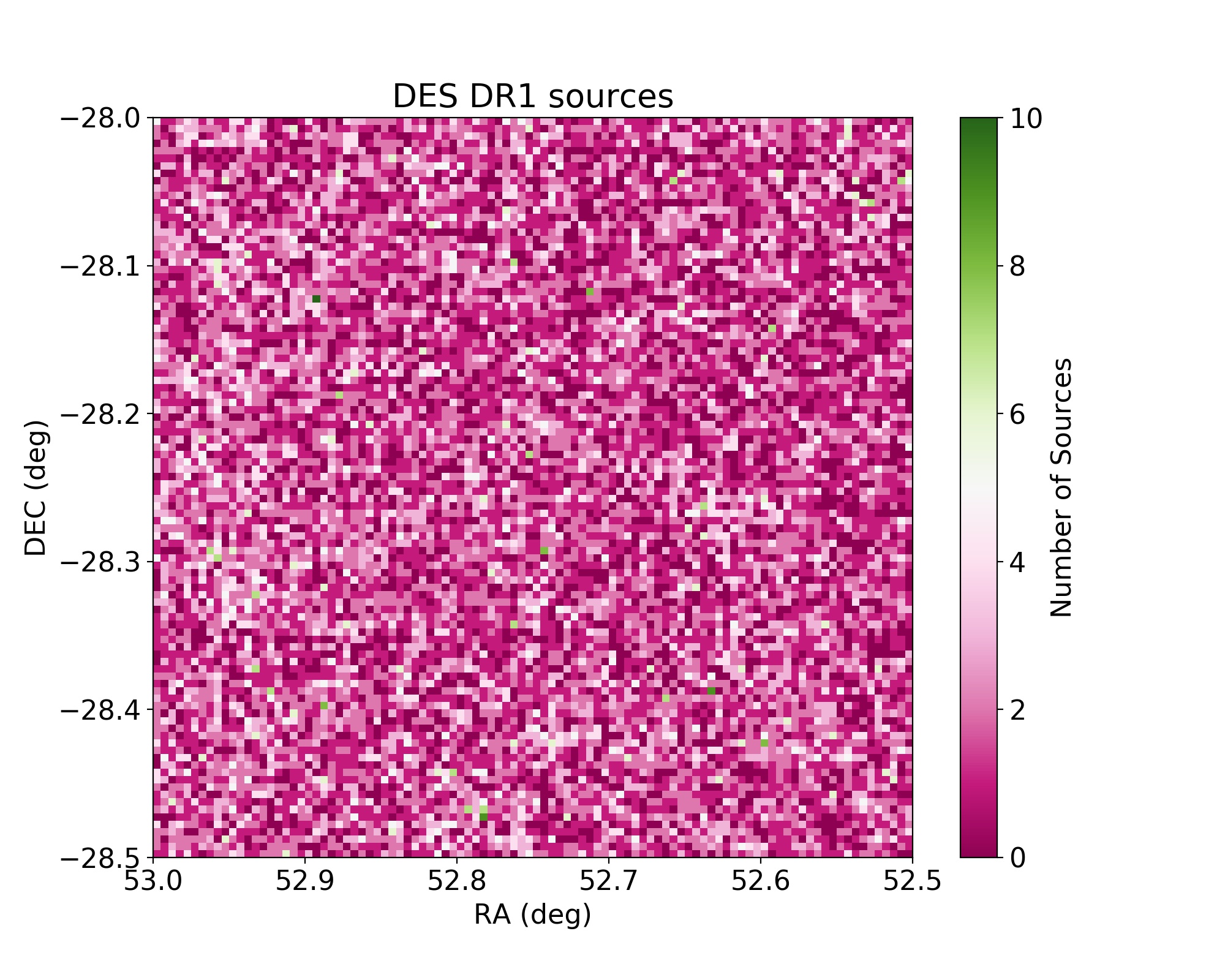}
   \includegraphics[scale=0.28]{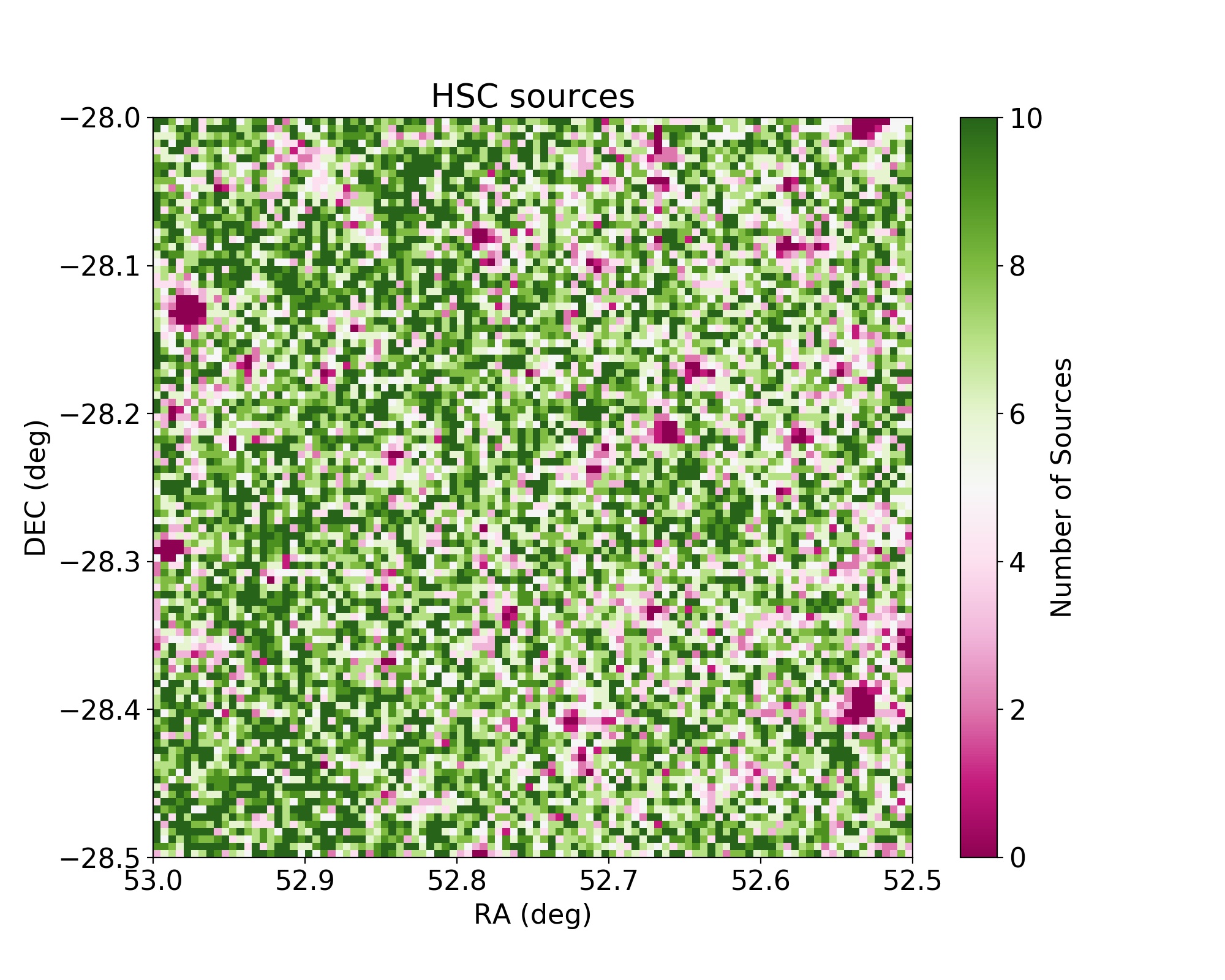}
   \caption{2D histograms of the number of $i$-band detected sources in DES DR1 (left) and HSC (right) in the same 0.25 $\rm deg^2$ region in W-CDF-S. Note that the two panels have the same color scale. The HSC observations increase the number density of optical sources by a factor of $\approx$ 10.}
\end{figure}

\renewcommand\thefigure{S5} 
\begin{figure}[htbp]
   \centering
   \includegraphics[scale=0.45]{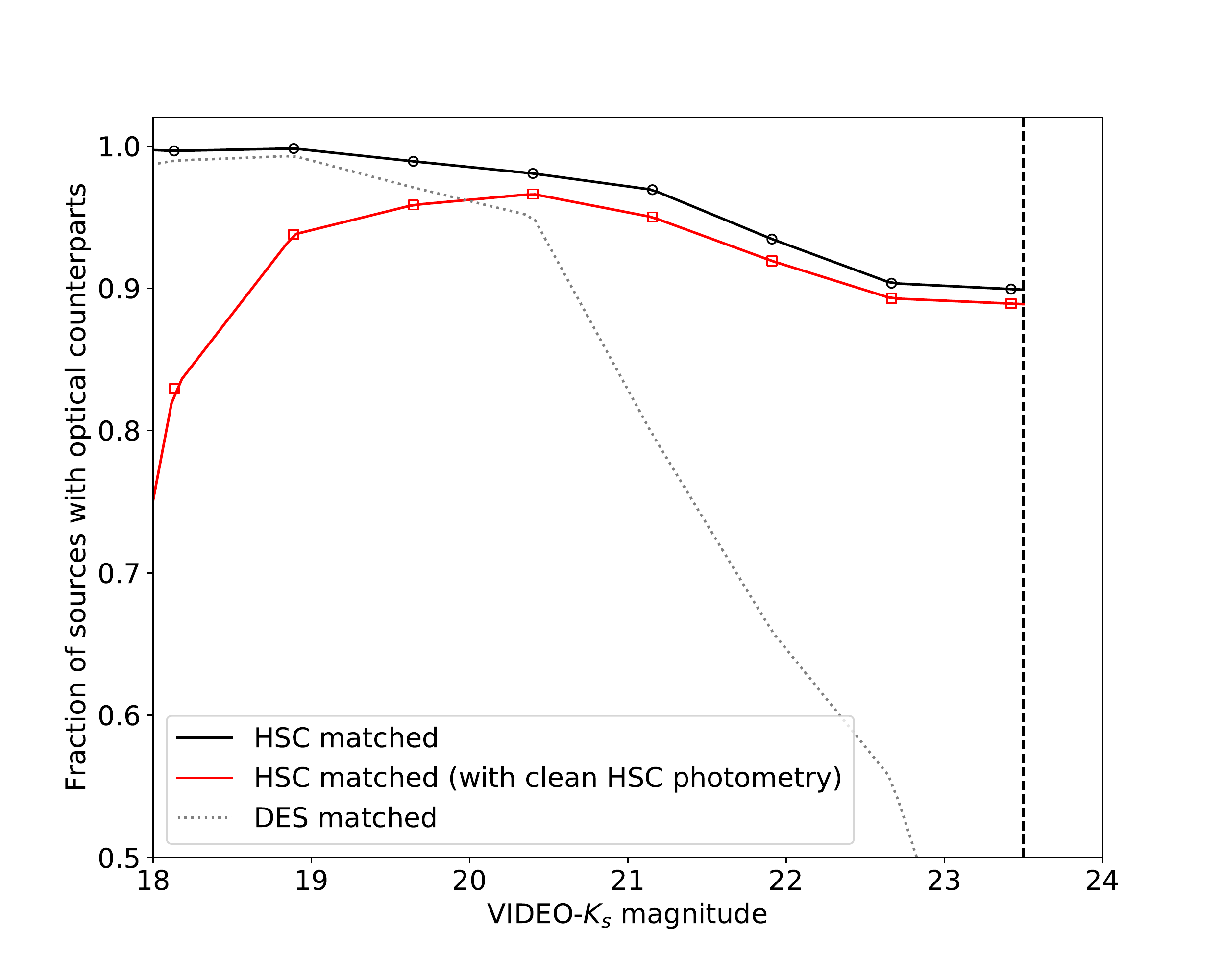}
   \caption{The fraction of VIDEO sources that have optical counterparts within 1'' at the given VIDEO-$K_s$ magnitude vs. the VIDEO-$K_s$ magnitude. 
   The black open circles and the black solid line represent the fraction of VIDEO sources with HSC counterparts at the given magnitude; the red open squares and the red solid line represent the fraction of VIDEO sources with ``clean'' HSC counterparts (\texttt{flag\_clean == True}) at the given magnitude; the gray dotted line represents the fraction of VIDEO sources with DES counterparts at the given magnitude. The vertical black dashed line represents the limiting magnitude of the $K_s$-band reported in VIDEO. 
   The HSC data are deep enough to match with VIDEO sources, while the DES data are not deep enough to match with VIDEO sources. The drop of the matched fraction with ``clean" HSC objects at the bright end is a result of the low saturation level of HSC (for ``clean" HSC objects, we require that the center of the source is not saturated).}
\end{figure}

\renewcommand\thefigure{S6} 
\begin{figure}[htbp]
   \centering
   \includegraphics[scale=0.45]{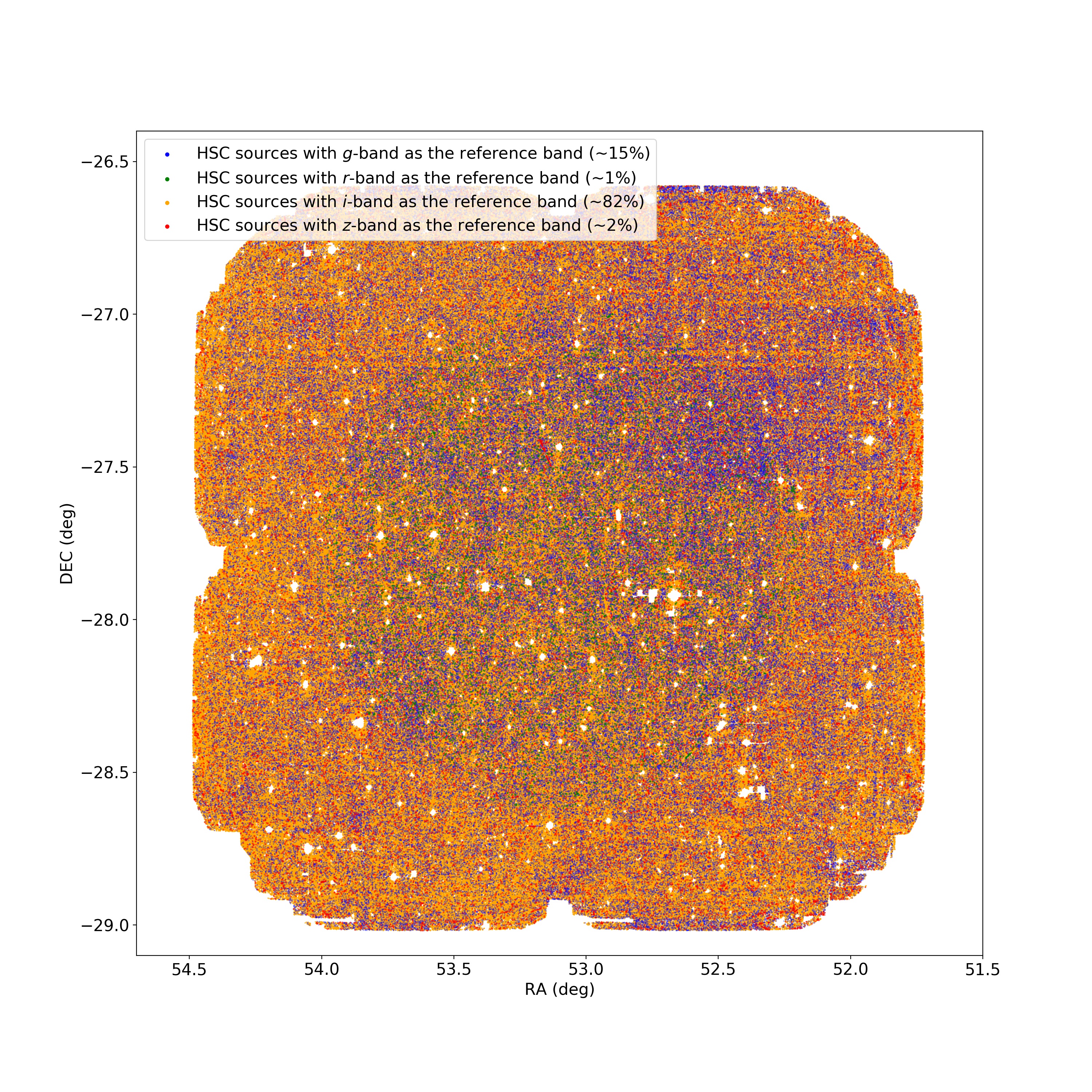}
   \caption{The map of ``clean" objects in the forced-photometry catalog. Circles with different colors represent sources with different reference bands used when performing the forced photometry. Most of the sources have the $i$-band as the reference band.}
\end{figure}

\renewcommand\thefigure{S7} 
\begin{figure}[htbp]
   \centering
   \includegraphics[scale=0.22]{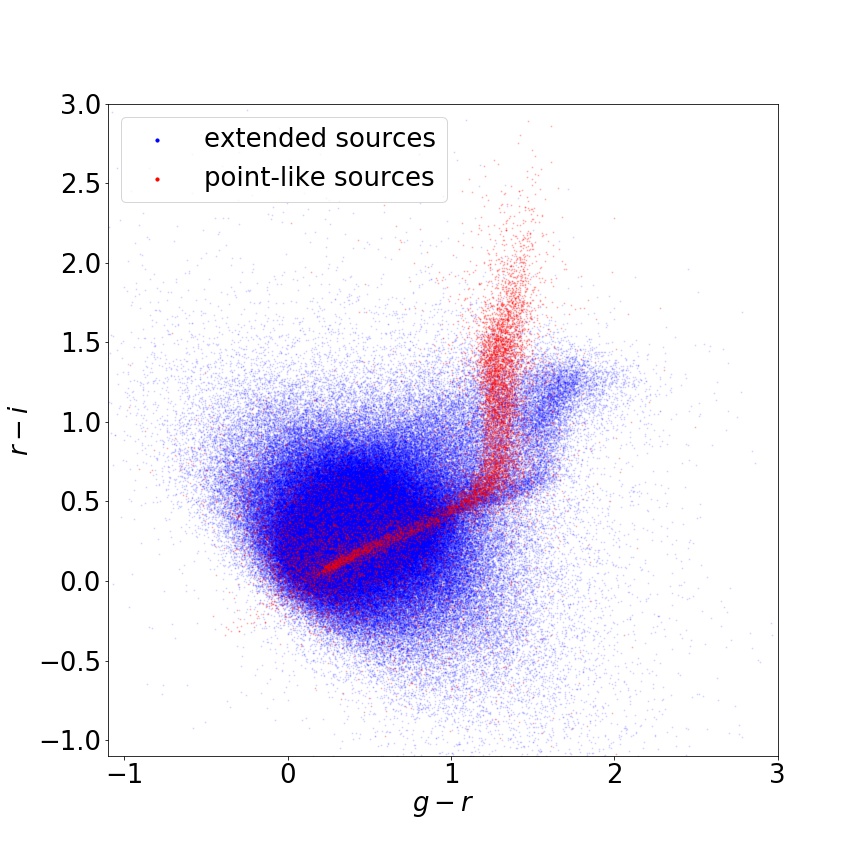}
   \includegraphics[scale=0.22]{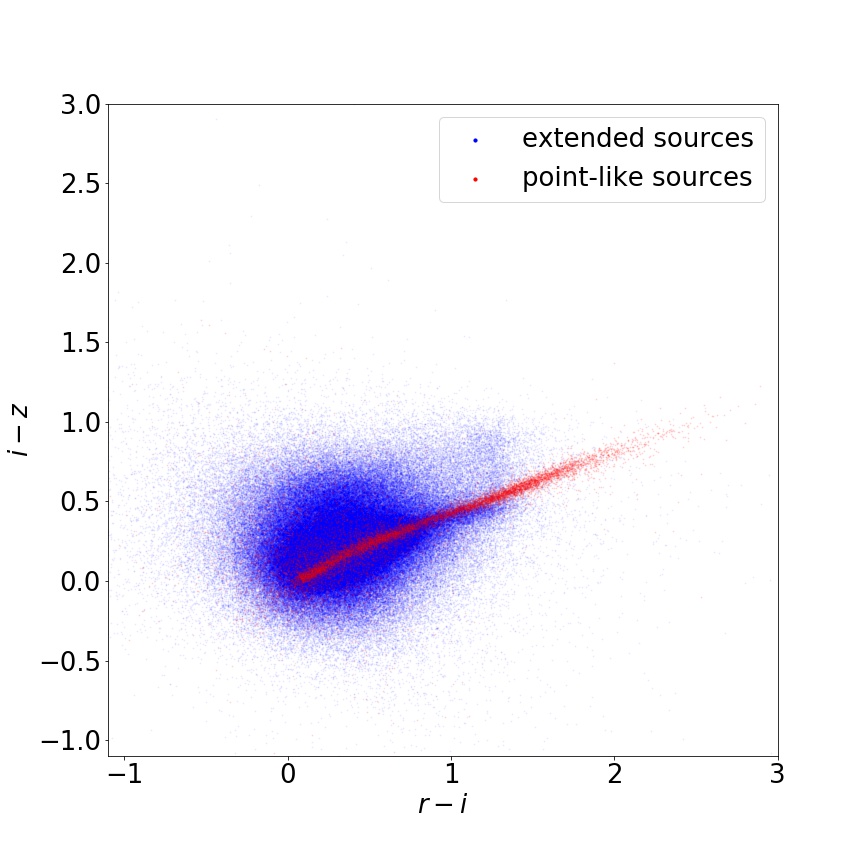}
   \caption{$g-r$ vs. $r-i$ (left) and $r-i$ vs. $i-z$ (right) plots for ``clean'' objects in the coadded HSC catalog with $i$-band magnitude $< 25$, and the magnitudes in the $g,i,z$ bands being brighter than the estimated depth of that band (the magnitudes have not been corrected for Galactic extinction). 
   The extended sources (represented by the blue dots) are selected by requiring \texttt{hsm\_point\_like == False}, 
   and the \hbox{point-like} sources (represented by the red dots) are selected by requiring \texttt{hsm\_point\_like == True}.}
\end{figure}

\end{document}